\input{aipcheck}

\documentclass[
    ,final            
  ]
  {aipproc}

\layoutstyle{6x9}

\begin{document}

\title{A New Class of Radio Pulsars - Back in 1982}


\author{M. Ali Alpar}{
 address={Sabanc{\i} University, Orhanl{\i}-Tuzla 34956, \.{I}stanbul, Turkey} }

\begin{abstract}
\\
\\
\emph{Dedicated to the memory of Jacob Shaham, who knew millisecond
X-ray pulsars would be discovered one day.}\\
\\
Basic ideas about the torques on the neutron star and the existence
of an equilibrium rotation period  followed from the recognition
that most X-ray binaries contain accretion powered neutron stars.
The evolution of binaries through a phase of accretion onto the
neutron star, eventually leading to a post- accretion radio pulsar
phase, was initially discussed as a way to understand the scarcity
of binaries among the radio pulsars and the relatively short
rotation periods of the first discovered binary radio pulsars in
terms of magnetic fields that would be smaller than the familiar
$10^{12} G$ range. The discovery of the millisecond pulsars  made us
realize that the fields can be much lower in a new class of radio
pulsars that have been spun up by accretion in LMXBs. The predicted
spin-down rates of the millisecond pulsar was soon confirmed. The
observers' search for millisecond X-ray periods was on, leading
first to the discovery of QPOs, and eventually to the discovery of
the X-ray millisecond pulsars. The theorists' quest for explanations
of why X-ray millisecond pulsations are not observed from LMXBs also
started right away.

\end{abstract}

\maketitle


\section{The Time Line}

  The time line of pulsar and neutron star observations starts from
  the discovery of the radio pulsars and proceeds with discoveries
  and important new timing observations every year or every few
  years, until a break between 1985 and 1998, ending with the
  discovery of the accreting millisecond pulsar. The
  discovery of radio pulsars (Hewish et al 1968) and the understanding of their basic nature
  (Pacini 1968, Gold 1968), the proposal of accretion powered neutron
  stars in X-ray binaries (Zeldovich \& Guseynov 1966) and the discovery of Sco X-1 (Giacconi et al. 1964), of
  pulsations from Cen X-3 (Giacconi et al 1971) and of Doppler shifts of the pulse period confirming the binary nature of
  this source (Schreier et al 1972) all followed one after the
  other, culminating with the introduction
  of the concepts of Alfven radius, corotation radius and rotational
  equilibrium (Pringle \& Rees 1972, Davidson \& Ostriker 1973, Lamb, Pethick \& Pines
  1973).\\

  Enquiring the future evolution of Her X-1, which was then, as it still is,
  one of the most interesting and most studied sources, together with the absence or relative
  scarcity of radio pulsars in binaries, led Bisnovatyi-Kogan \& Komberg to propose that accreting
  neutron stars in binaries have weak magnetic fields, giving rise, at the end of the accretion phase,
  to low radio luminosity pulsars which are difficult to detect (Bisnovatyi-Kogan \& Komberg 1974).
  According to these authors accretion would lead not only to spin up, but also to suppression of the magnetic field.
  Accreting throughout the long lifetime of a Her X-1 type X-ray progenitor system, the neutron star would end up
  with $B \sim 10^8-10^9 \;G$, small enough for lack of radio pulsar activity post accretion.
  Many of the important ideas concerning the evolution of LMXBs to yield low magnetic field,
  short rotation period neutron stars are already present in the work of Bisnovatyi-Kogan \& Komberg, \emph{except that
  the endpoint is not a radio pulsar} because they do not consider the possibility that the neutron star rotation period
  could become short enough to make an active radio pulsar in spite of the low $B$.\\

  The first radio pulsar in a binary, PSR B 1913+16, was discovered in the following year (Hulse \& Taylor
  1975). The short (59 ms) rotation period of PSR B 1913+16
  was explained by Smarr \& Blandford (1976) as the equilibrium period reached in a previous accretion
  phase: a short equilibrium period would correspond to a weak
  magnetic field. The relatively weak $10^{11} G$ magnetic field of the pulsar in the Hulse-Taylor binary could itself be
  attributed to its accretion history. This idea was elaborated by Backus, Taylor and Damashek (1982), who applied it to the
  binary radio pulsars then known, and noted that the binary pulsars were all located in the lower left corner of the
  $P-\dot{P}$ diagram, somewhat off from the distribution of most radio pulsars. They further speculated
  that the isolated radio pulsar PSR B 1952+29, whose location in the $P-\dot{P}$ diagram is close to the positions
  of the binary pulsars PSR B 1913+16 and PSR B 0655+64, had also been spun up by accretion in a binary
  which was subsequently disrupted by a second supernova explosion, of the companion donor star.\\

  The discovery of an unexpectedly short period pulsar, the first millisecond pulsar PSR B 1937+21
  (Backer, Kulkarni, Heiles, Davies \& Goss 1982) was then immediately explained as the result of accretion
  in low mass X-ray binaries (LMXBs) by two groups independently (Alpar, Cheng, Ruderman \& Shaham
  1982; Radhakrishnan \& Srinivasan 1982). Both groups made the bold inference that the LMXB magnetic fields
  must be as weak as $10^8-10^9 \;G$ in order to lead to millisecond equilibrium rotation
  rates, and predicted that the spin-down rate of the millisecond
  radio pulsar would be as low as $10^{-19} s \;s^{-1}$.\\

  In the Alpar et al. paper we posed the argument starting from properties of LMXBs,
  in particular the expectation that the neutron stars in these systems have weak magnetic fields.
  The magnetic field and the mass accretion rate in the LMXB phase
  determine the equilibrium period, which would be the rotation period of the radio pulsar at its birth.
  The subsequent spin-down rate is related to the magnetic field and rotation period through the dipole radiation
  relation. Eliminating the magnetic field between these two relations gives the "spin-up" or "birth" line
  shown for the first time in the $P-\dot{P}$ diagram in our paper. The first millisecond pulsar PSR B 1937+21,
  lacking a proper pulsar name yet, is indicated as 4C21.53 below the birth line in the far bottom left
  corner of the $P-\dot{P}$ diagram, now extended by several decades in $P$ and $\dot{P}$. All the binary radio pulsars
  as well as the isolated pulsar PSR B 1952+29 are below the spin-up line. Depending on the time averaged
  mass accretion rate, which we assumed to be typically $0.1 \dot{M}_{Edd}$, the millisecond
  \emph{radio} pulsar evolves on a dipole spin-down track proceeding
  from the birth line. The part of the $P-\dot{P}$ diagram populated by the millisecond and binary pulsars
  lies in the region of radio pulsar activity, clearly
  above the so called "pulsar death valley".
  We thus predicted that the spin-down rate of the
  millisecond pulsars should be in the $10^{-19} s \;s^{-1}$ range or less. This prediction was soon verified by the
  measurement of the spin-down
  rate of PSR B 1937+21, as $\dot{P} \cong 1.2 \times 10^{-19} s \;s^{-1}$ (Backer, Kulkarni \& Taylor 1983).
  We also noted that if the dipole magnetic field of a millisecond
  pulsar were in the $10^{12} \;G$ range typical of the conventional radio pulsars, the pulsar magnetosphere would
  sustain voltages $10^6$ times higher than that of the Vela pulsar, and would therefore power high energy radiation
  which is not observed. Such a pulsar would also spin down very rapidly,
  after only a brief epoch in the millisecond period range. \\

  Radhakrishnan \& Srinivasan (1982) started their argument by noting the
  lack of a supernova remnant, or any X-ray emission from a nebula that should
  be powered by a necessarily young millisecond pulsar, if it had a
  conventional magnetic field. They used
  the observational upper limits on the x-ray luminosity of the
  source to deduce, empirically, that the dipole magnetic field must
  be less than about $4 \times 10^8 G$ and the spin down rate must
  be less than about $10^{-19} s s^{-1}$. They then proceeded to
  note that such a weak magnetic field would yield spin-up to a
  millisecond rotation period as the equilibrium period after
  accretion in a binary system. After these first two papers a third,
  independent paper by Fabian, Pringle, Verbunt\& Wade (1983) proposed the same
  evolutionary scenario for the millisecond pulsar, without making a prediction for the spin-down
  rate. The measurement by Backer, Kulkarni \& Taylor (1983) was published a week after this
  paper.\\

  Since the millisecond radio pulsar was supposed to have been spun up in a LMXB,
  some LMXBs should have millisecond rotation periods.
  Thus the search for accreting millisecond X-ray pulsars started in 1982.
  The search first yielded quasi-periodic oscillations rather than the expected pulses
  at the exact rotation period. The first QPOs were discovered in 1985 from the LMXB GX 5-1
  (van der Klis et al 1985). These QPOs had frequencies $\nu\sim $ 20-40 Hz which varied in
  correlation with the count rate. The explanation, provided by Alpar \& Shaham (1985), was the beat frequency
  model. According to this model, the observed QPO frequencies are beat (difference) frequencies,
  between keplerean rotation rates in the inner edge of the accretion disk and the neutron star's rotation rate.
  Asymmetric structures in the star's magnetosphere and in the disk match periodically at the beat
  frequency, thus superimposing a modulation of the mass accretion rate (and the observed count rate) at the beat
  frequency. Evaluating the observed correlation between the QPO frequency and the count rate in terms of the model,
  Alpar \& Shaham estimated the rotation period GX 5-1 to be 6.6 ms and the magnetic field to be
  $5.5 \times 10^9 G$. This indirectly confirmed the LMXB-millisecond pulsar connection.\\

  The every-few-years rhythm of pulsar timing discoveries continued from 1967 to 1985. The
  time line then goes into a different mood, a lull decorated with a
  rich set of discoveries of more and different types of QPOs (van der Klis 2006), which lasted until the discovery of
  millisecond X-ray pulsations from the LMXB SAX 1808.4-3658 by Wijnands \& van der Klis (1998).

\section{Why is it difficult to observe an accreting millisecond X-ray pulsar?}

  The lack of observations of accreting millisecond x-ray pulsars
  preoccupied astrophysicists until the discovery by Wijnands \& van der Klis
  in 1998; to understand the scarcity of these sources continues to be of
  interest.\\

  The question of the observability of accreting millisecond X-ray
  pulsars was raised in our first paper, where we noted that "Certainly its among
  the galactic bulge X-ray sources that
  millisecond neutron star periods are common" and also that "in such neutron stars there is no sign of the
  modulation of X-ray emission expected from rotation of a magnetized stellar surface with $B >> 10^{10} \;G$ ",
  invoking this to argue
  for weaker magnetic fields and thereby implying that the weak field, which would be
  relatively inefficient in channeling the accretion flow, is the reason for
  the lack of X-ray pulsations from LMXBs
  (Alpar et al 1982). Fabian et al (1983) noted that "The lack of
  observed short pulse periods in the galactic bulge sources
  indicates either that the fields are weak enough not to disrupt
  the disk and so to produce an observable modulation or that the
  periods are short enough to be smeared out by the electron
  scattering region which appears to surround at least some of the
  sources."\\

  The question was revived by the discovery of the first
  QPOs (van der Klis et al. 1985). The beat frequency model requires
  the magnetic field of the neutron star to be strong enough to actually
  effect the channeling of the accretion flow, so that the weakness of the
  field could not be the explanation for the lack of observed X-ray pulsations
  at the neutron star rotation period. A different reason is noted in the beat frequency paper by Alpar
  \& Shaham (1985), "that obscuration and inobservability
  of the fundamental stellar pulsations at $\Omega_0$ may be here
  simply due to a thick disk". The follow-up paper (Lamb, Shibazaki,
  Alpar \& Shaham 1985) listed the possible reasons why
  an X-ray signal pulsed at the rotation rate of the neutron star
  might not be produced or might not emerge from the source.
  Three classes of mechanisms are still being discussed:\\

  a) Weak fields or weak beaming near the neutron star surface: While the dipole magnetic field is strong enough to stop
  the disk at $r_A > R_{\*}$, it is not strong enough to channel the accretion
  significantly to produce a beamed X-ray signal so that the modulation at the
  rotation period is too broad and has low pulsed amplitude. Spectral evidence from some LMXBs indicating that
  accretion is spread over a large fraction of the surface supports this hypothesis, subject to ambiguities in
  interpretation of the spectra. Gravitational bending of radiation,
  higher multipole structure of the field near the neutron star
  surface and radiative transfer through the accretion column leading to broad,
  fan-like radiation patterns are further mechanisms that may
  suppress beaming at the source.
  The peak in the power spectrum at the rotation frequency  may be
  broadened due to sidebands from fluctuations in the accretion flow/fluctuations
  in X-ray brightness, or due to the motion of bright spots on the neutron star surface.
  We now know this happens, as evidenced by the burst oscillations.\\

  b) Obscuration and selection effects: Geometrically and optically
  thick disks and disk coronae allow us to see X-rays from the
  neutron star surface only for systems viewed at small inclination with the disk
  (and neutron star) rotation axis; hence observed systems have weak rotational modulation.
  This explanation is possible for some sources; its general applicability depends on
  how common sources with thick disks are.\\

  c) Compton scattering: A Compton scattering corona with electron scattering optical
  thickness $\tau_{es} \sim 1$ surrounding the neutron star destroys
  the \emph{beaming} of the X-rays, so the signal carrying the neutron star rotation period is
  lost. Note that the criterion for loss of beaming due to Compton
  scattering is much more stringent than the criterion for the
  smearing of periodicity or time dependence of an isotropic
  radiation field. Beaming is lost at $\tau_{es} \sim 1$, as one
  scattering is enough to randomize the direction of the photon
  momentum, whereas modulation at a timescale $\Delta t$ in isotropic radiation will be smeared out in a
  comptonizing medium of size $R$ only if
  $\tau_{es} R / c > \Delta t$, which will typically require large
  $\tau_{es}$. Thus a source embedded in a comptonizing corona would
  kill a beamed signal carrying the neutron star rotation period
  while allowing QPO modulation at periods comparable to the
  rotation to be observed.\\

  A recent check on this idea (G\"{o}\u{g}\"{u}\c{s}, Alpar \& Gilfanov 2007) showed that for at least
  some LMXBs the destruction of beaming by Compton scattering
  is not the likely explanation for the lack of pulsations at the rotation period.
  The spectra of the sources investigated,
  GX 9+1, GX 9+9 and Sco X-1 indicate that for reasonable electron
  temperatures in the corona $\tau_{es}$ is significantly less than 1.
  Thus the radiation is not comptonized, and the reason for the suppression of pulses at
  the rotation period must be something other than comptonization.\\

  On the other hand there are some accreting X-ray millisecond pulsars with $\tau_{es} >
  1$. Indeed the pulsed component itself displays a comptonized spectrum, as discussed
  in this meeting (Falanga 2008, Poutanen 2008).
  It is easy to understand this in terms of geometry. If the comptonizing corona is localized on the pulsar's polar caps
  or other regions at or near the surface where the X-ray beam is produced,
  loss of beaming would lead to comptonized and isotropized
  radiation. Since this comes only from some patches, "hot spots", on the neutron star's
  surface, a comptonized signal modulated at the rotation period emerges.
  \\
  \\
  \\
  \\
  \\
  \\
  \\
  \\
\begin{theacknowledgments}
 I acknowledge research support from the Sabanc{\i} University Astrophysics \& Space Forum, from the EC FP6 Marie Curie
 Transfer of Knowledge Project ASTRONS and from the Turkish Academy of Sciences. Participation in this conference was supported by the Turkish
 Academy of Sciences.
\end{theacknowledgments}



\end{document}